\documentclass[aps,prb,twocolumn,notitlepage,groupedaddress,showpacs]{revtex4-1}

\usepackage{graphicx}

\begin{document}


\title{Coherence and stiffness of spin waves in diluted ferromagnets}


\author{I. Turek}
\email[]{turek@ipm.cz}
\affiliation{Institute of Physics of Materials,
Academy of Sciences of the Czech Republic,
\v{Z}i\v{z}kova 22, CZ-616 62 Brno, Czech Republic}

\author{J. Kudrnovsk\'y}
\email[]{kudrnov@fzu.cz}
\affiliation{Institute of Physics, 
Academy of Sciences of the Czech Republic,
Na Slovance 2, CZ-182 21 Praha 8, Czech Republic}

\author{V. Drchal}
\email[]{drchal@fzu.cz}
\affiliation{Institute of Physics, 
Academy of Sciences of the Czech Republic,
Na Slovance 2, CZ-182 21 Praha 8, Czech Republic}


\date{\today}

\begin{abstract}
We present results of a numerical analysis of magnon spectra in
supercells simulating two-dimensional and bulk random diluted
ferromagnets with long-ranged pair exchange interactions.
We show that low-energy spectral regions for these strongly
disordered systems contain a coherent component leading to
interference phenomena manifested by a pronounced sensitivity of
the lowest excitation energies to the adopted boundary conditions.
The dependence of configuration averages of these excitation
energies on the supercell size can be used for an efficient
determination of the spin-wave stiffness $D$.
The developed formalism is applied to the ferromagnetic Mn-doped
GaAs semiconductor with optional incorporation of phosphorus;
the obtained concentration trends of $D$ are found in reasonable
agreement with recent experiments.
Moreover, a relation of the spin stiffness to the Curie temperature
$T_\mathrm{C}$ has been studied for Mn-doped GaAs and GaN
semiconductors.
It is found that the ratio $T_\mathrm{C}/D$ exhibits qualitatively
the same dependence on Mn concentration in both systems.
\end{abstract}


\maketitle


\section{Introduction\label{s_intr}}

Magnon spectra of ferromagnetic systems represent a subject
of long-lasting theoretical and experimental research, motivated
by the well-known relevance of magnons for finite-temperature
properties and for the magnetization dynamics on a microscopic
scale. 
Qualitative and quantitative features of the magnon spectra
become especially important for recently investigated low-dimensional
systems, such as ultrathin films and nanowires, as well as for
disordered or diluted bulk systems, such as metallic solid solutions
and diluted magnetic semiconductors.

On the theoretical side, the simplest approaches to the magnon
spectra employ various Heisenberg Hamiltonians with pair exchange
interactions between localized spins.
The resulting equations of motion for spin excitations are solved by
means of the Tyablikov decoupling procedure (random-phase
approximation, RPA) \cite{r_1967_svt, r_2000_nm_b} which yields an
improved description as compared to the usual mean-field
approximation (MFA).
For finite-temperature properties, other methods of statistical
physics including Monte Carlo simulations can be used as well.

The treatment of magnons in random ferromagnets with crystalline
structures is obviously more difficult than in the case of perfect
nonrandom crystals since the spin waves propagate through a random
system even at zero temperature.
In analogy with realistic models of lattice vibrations (phonons),
the disorder in the effective Hamiltonian for the magnon dynamics,
i.e., the dependence of the Hamiltonian matrix elements on the
occupation of lattice sites by various atomic species of the random
alloy, is quite complicated as a consequence of the Goldstone sum
rule \cite{r_2000_ag}.
This type of disorder makes the formulation and numerical
implementation of an effective-medium theory for magnons rather
demanding as documented, e.g., by recent studies \cite{r_2002_bb,
r_2016_bsb}. 
For these reasons, brute-force numerical approaches have often 
been used for magnon spectra and critical behavior of
disordered ferromagnets, ranging from Monte Carlo
simulations \cite{r_2016_bsb, r_2004_bek, r_2004_ssd} over a
direct solution of the spin dynamics equations \cite{r_2008_shn}
to a semi-analytic real-space RPA \cite{r_2004_hn, r_2005_bzk_e,
r_2007_gb}.

These approaches have been quite successful in quantitative
description of diluted magnetic semiconductors, such as
Mn-doped GaAs \cite{r_2010_sbk}.
As it has been emphasized by several authors, this diluted
system is featured by a strong disorder which has a profound
effect on its basic quantities, such as the Curie temperature
$T_\mathrm{C}$, see Ref.~\onlinecite{r_2004_bek, r_2004_ssd,
r_2005_bzk_e}, and the spin-wave stiffness $D$, see 
Ref.~\onlinecite{r_2007_gb}.
In particular, the high degree of dilution (content of Mn atoms
below 10 at.~\%) makes effects of local fluctuations very important.
Depending on the Mn concentration and the spatial extent of the
pair exchange interactions, percolation phenomena become important
as well for Mn-doped GaAs and similar diluted ferromagnets.
As a consequence, simple theoretical techniques, such as the MFA
or the virtual-crystal approximation (VCA), fail to provide reliable
values of $T_\mathrm{C}$, $D$ and other physical quantities
\cite{r_2010_sbk}.

However, the very strong disorder cannot suppress all signs of
phase coherence of the spin waves in Mn-doped GaAs.
This feature has been observed and used in recent magneto-optical
pump-and-probe experiments on (Ga,Mn)As epitaxial thin films
\cite{r_2013_nnt, r_2015_srt}, in which the stiffness $D$ has been
determined from frequencies of spin-wave resonances. 
This procedure rests on the original theory developed by Kittel for
microscopically homogeneous ferromagnetic thin films with specific
boundary conditions \cite{r_1958_ck}.
In the theoretical studies of Mn-doped GaAs, the spin-wave stiffness
$D$ has been derived from positions of peaks in the energy- and
momentum-resolved magnon spectral function \cite{r_2016_bsb,
r_2007_gb}; the employed periodic boundary conditions play only
a technical role in these approaches.
Let us note that a coherent component has also been found in the
spectra of other excitations in disordered media, such as, e.g.,
acoustic waves in granular systems \cite{r_1999_jcv, r_2005_srs}.

The purpose of this paper is to examine theoretically spectral
properties of spin waves of strongly disordered diluted ferromagnets
in the long-wavelength limit from a point of view of the phase
coherence. 
We find that this approach enables one to determine the spin-wave
stiffness $D$ without an intermediate calculation of the energy-
and momentum-resolved spectral function.
As an application, the values of $D$ are studied for (Ga,Mn)As,
(Ga,Mn)(As,P) and (Ga,Mn)N diluted magnetic semiconductors in
order to answer some actual questions, such as the effect of
doping by phosphorus on micromagnetic parameters of (Ga,Mn)As
\cite{r_2015_srt, r_2014_tbc} or the relation between the spin
stiffness and the Curie temperature \cite{r_2010_wd}.

\section{Models and methods\label{s_meth}}

Our formalism employs the isotropic Heisenberg model for classical
magnetic moments of the same fixed magnitude $M$ located at lattice
sites labelled by indices $m, n, r$.
The energy of each configuration $\{ \mathbf{e}_m \}$, where
the unit vector $\mathbf{e}_m$ defines the direction of the
local moment on site $m$, is given by
\begin{equation}
\mathcal{E} = - \sum_{mn} J_{mn} \mathbf{e}_m \cdot \mathbf{e}_n ,
\label{eq_chm}
\end{equation}
where the quantities $J_{mn}$ denote the pair exchange interactions.
They satisfy usual properties ($J_{mm} = 0$, $J_{mn} = J_{nm}$) and
are mostly nonnegative, so that the ground state is ferromagnetic.
In the Green's function formalism \cite{r_1967_svt, r_2000_nm_b,
r_2004_hn, r_2005_rtd, r_2005_ssb}, the zero-temperature spin-wave
propagator is (apart from a prefactor) equivalent to the resolvent
of the Hamiltonian $H$ with matrix elements
\begin{equation}
H_{mn} = \delta_{mn} \left( \sum_r J_{mr} \right) - J_{mn} 
\label{eq_hme}
\end{equation}
and the energies of spin excitations are directly related to the
eigenvalues of $H$.
In the special case of a perfect crystalline ferromagnet on a
Bravais lattice, the magnon energy for a reciprocal-space vector
$\mathbf{q}$ can be obtained as
\begin{equation}
E^\mathrm{mag}(\mathbf{q}) = \frac{4 \mu_\mathrm{B}}{M}
\left[ J(\mathbf{0}) - J(\mathbf{q}) \right] ,
\label{eq_magen}
\end{equation}
where $\mu_\mathrm{B}$ is the Bohr magneton and the quantity
$J(\mathbf{q})$ denotes the lattice Fourier transform of the
translationally invariant pair exchange interactions $J_{mn}$.
For lattices with a high-symmetry point group, such as the square
lattice in two dimensions or fcc and bcc lattices in three
dimensions, the magnon energy for small wave vectors $\mathbf{q}$
follows a quadratic law
\begin{equation}
E^\mathrm{mag}(\mathbf{q}) \approx Dq^2 ,
\label{eq_dsw}
\end{equation}
where $D$ is the spin-wave stiffness and $q = |\mathbf{q}|$.

The diluted ferromagnets considered in this study employ nonrandom
translationally invariant pair exchange interactions $J_{mn}$ defined
on a Bravais lattice, but with the local moments occupying randomly
only a part of all lattice sites.
The concentration of the magnetic sites is denoted by $x$ 
($0 < x < 1$).
We studied two different cases, both corresponding to long-ranged
exchange interactions and a strong dilution ($x \ll 1$); the models
were treated on large supercells with periodic boundary conditions.
The individual configurations were constructed by means of a
random-number generator and the eigenvalues of the
Hamiltonian~(\ref{eq_hme}) were found numerically
by means of standard techniques for the diagonalization
of real symmetric matrices \cite{r_1965_jhw, r_1992_ptv}.

In the first case, a two-dimensional square lattice was chosen
with a model exchange interaction
\begin{equation}
J_{mn} = A (a/d_{mn})^2 \exp(-d_{mn}/\lambda) , 
\label{eq_j2d}
\end{equation}
where $A$ is the interaction strength, $a$ denotes the lattice
parameter, $d_{mn}$ is the distance between lattice sites $m$
and $n$, and $\lambda$ denotes a screening length.
The inverse proportionality to the square of the distance
in the model~(\ref{eq_j2d}) coincides with the non-oscillating
part of the Ruderman-Kittel-Kasuya-Yosida interaction in
two-dimensional metals \cite{r_1975_fk} while the exponential decay
with increasing distance reflects disorder-induced damping of the
interaction.
We choose $\lambda = 5a$ and neglected all
interactions~(\ref{eq_j2d}) for $d_{mn} > 35a$.
Supercells of two shapes were constructed with an integer $L$
controlling their size: (i) squares with edges $La$ comprising $L^2$
sites ($100 \le L \le 300$), and (ii) rectangles with edges $La$ and
$2La$ comprising $2L^2$ sites ($75 \le L \le 180$).
The concentration of local moments was chosen $x = 0.03$; this
small concentration is still above the percolation threshold owing
to the big values of the screening length $\lambda$ and the
cut-off distance of the exchange interaction.

In the second case, realistic models of bulk Mn-doped GaAs were
constructed with magnetic Mn atoms occupying randomly the
fcc Ga-sublattice of the zinc-blende structure of GaAs.
The Mn-Mn exchange interactions were derived for each Mn
concentration $x$ ($x < 0.1$) from the \emph{ab initio} electronic
structure \cite{r_2004_ktd_p} using the magnetic force theorem
\cite{r_1987_lka}; the cut-off distance was taken as $6.36 a$,
where $a$ denotes the lattice parameter of the zinc-blende
structure.
Supercells of two kinds were treated: (i) cubic supercells with
edges $La$ and with $4L^3$ fcc-lattice sites for Mn atoms
($18 \le L \le 38$),
and (ii) rhombohedral supercells with edges given by three vectors
$(La/2 , La/2 , 0)$, $(La/2 , 0 , La/2)$ and $(0 , La/2 , La/2)$,
which contain $L^3$ sites ($34 \le L \le 60$).

Reliable values of physical quantities follow from the supercell
approach after averaging over $N_\mathrm{conf}$ random
configurations; in agreement with previous studies \cite{r_2005_bzk_e,
r_2007_gb} we found that $N_\mathrm{conf} = 50$ provides sufficient
statistics in most cases.
For the sake of comparison, results of the simple VCA will
also be mentioned in the following.
In the studied models, they are obtained by replacing the original
pair interactions $J_{mn}$ by concentration-weighted effective
interactions $xJ_{mn}$ (and by occupying the whole Bravais
lattice by the local moments).
As an example, we get from (\ref{eq_magen}) a relation
\begin{equation}
E^\mathrm{mag}_\mathrm{VCA}(\mathbf{q}) = \frac{4 \mu_\mathrm{B}}{M} 
\, x \left[ J(\mathbf{0}) - J(\mathbf{q}) \right] 
\label{eq_magen_vca}
\end{equation}
representing the magnon energy in the VCA.

\section{Results and discussion\label{s_redi}}

\subsection{Two-dimensional model\label{ss_2dm}}

\begin{figure}
\includegraphics[width=0.45\textwidth]{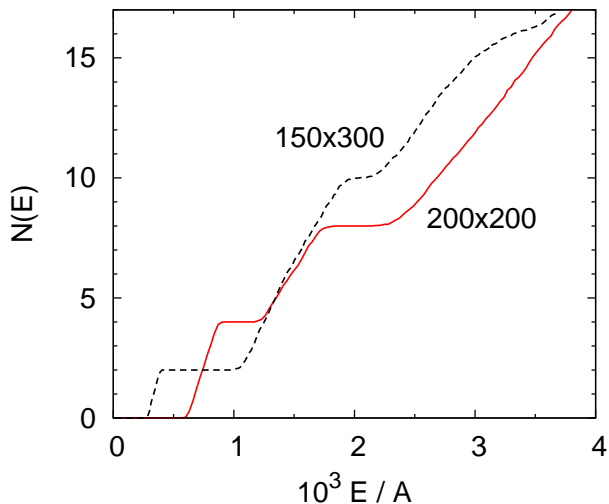}
\caption{%
The number of states $N(E)$ for the diluted 2D ferromagnet with
3\% of magnetic sites simulated by a square supercell with
$200\times 200$ sites (full line) and by a rectangular supercell
with $150\times 300$ sites (dashed line).
\label{f_ne2d}}
\end{figure}

The distribution of eigenvalues $E$ of the Hamiltonian~(\ref{eq_hme})
for the two-dimensional model can be inferred from
Fig.~\ref{f_ne2d}, which displays the low-energy tail of the
number of states function (integrated density of states) $N(E)$
averaged over 100 random configurations.
The zero eigenvalue, present in each configuration due to general
properties of the isotropic Heisenberg model~(\ref{eq_chm}), was
omitted in the calculation of $N(E)$.
The resulting $N(E)$ are featured by plateaux for supercells of
both shapes (square, rectangular); these plateaux correspond to an
absence of eigenvalues in certain energy intervals, which allows
one to separate the first lowest excited levels ($E_1$) from
higher-lying excitations.
The constant values of $N(E)$ in the plateaux are integers, which
point to four and two lowest excitations for the square and
rectangular supercells, respectively.

\begin{figure}
\includegraphics[width=0.45\textwidth]{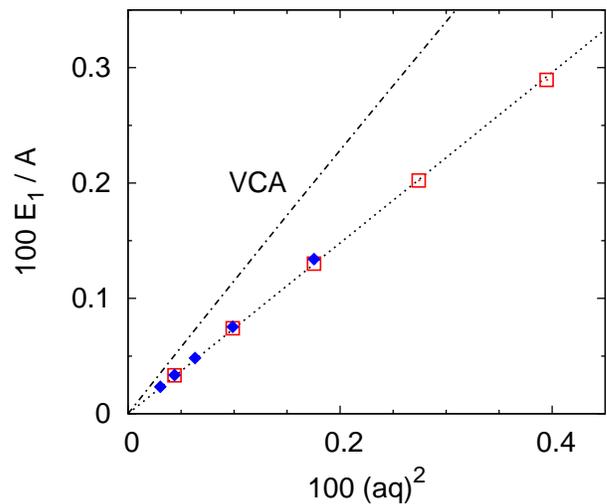}
\caption{%
The configurationally averaged
lowest excitation energies $E_1$ for the diluted 2D ferromagnet
as functions of the $\mathbf{q}$-vector magnitude derived from the
supercell size: for square supercells (open boxes) and for
rectangular supercells (full diamonds).
The straight dotted line results from a fit and
the dash-dotted line refers to the VCA-dependence.
\label{f_eq2d}}
\end{figure}

Based on an analogy with spin waves in nonrandom ferromagnets
treated with periodic boundary conditions \cite{r_1958_ck}, the
lowest excitations $E_1$ can be ascribed to the shortest
reciprocal-space vectors $\mathbf{q}$ compatible with the supercell
shape and size.
This identification yields four lowest excitations $E_1$ derived
from the vectors given by $\mathbf{q} = 2\pi (\pm 1 , 0) / (La)$ and
$\mathbf{q} = 2\pi (0 , \pm 1) / (La)$ for the square supercells,
but only two lowest excitations are obtained from $\mathbf{q} = \pi
(0, \pm 1) / (La)$ for the rectangular supercells employed.
These integers coincide with the constant $N(E)$ values inside the
plateaux, see Fig.~\ref{f_ne2d}.

This interpretation of the plateaux in $N(E)$ is further
corroborated by the dependence of the configurationally averaged
value of $E_1$ on the shortest $\mathbf{q}$-vector magnitude, as
obtained from variations of the supercell size $L$ and plotted in
Fig.~\ref{f_eq2d}.
One can see a proportionality relation $E_1 = k_1 q^2$ for
both shapes, with a common slope $k_1$ equivalent to the spin-wave
stiffness $D$ (note that $D/k_1 = 4\mu_\mathrm{B}/M$). 
The stiffness from the supercell calculations is significantly
reduced as compared to its VCA counterpart (Fig.~\ref{f_eq2d}),
in qualitative agreement with recent results for a diluted
two-dimensional ferromagnet \cite{r_2015_cws} with a slightly
different distance-dependence of the pair exchange interaction
($J_{mn} \propto d_{mn}^{-3}$).
A similar overestimation of $D$ in the VCA has also been observed
for the exchange interaction restricted to the first nearest
neighbors but with a much higher concentration of magnetic
atoms \cite{r_2016_bsb}.
These facts can be understood in terms of the lowest moments of the
$\mathbf{q}$-dependent magnon spectral function: its mean value,
given by the first moment and identical with the excitation energy
in the VCA, is comparable with the linewidth (the standard deviation
of the spectral function), given by the second moment, see
Ref.~\onlinecite{r_2007_gb} for details.

The results shown in Fig.~\ref{f_ne2d} and \ref{f_eq2d}
prove that the dispersion law of low-energy spin
excitations can easily be obtained only from the sensitivity of
the lowest eigenvalues of the Hamiltonian~(\ref{eq_hme}) to
adopted boundary conditions of the simulation supercells.
This sensitivity reflects the coherent part of the magnon
spectrum, which is present despite the strong disorder and
dilution. 
Theoretically, the presence of the coherent part rests on a very
weak damping of low-energy magnons, see
Ref.~\onlinecite{r_2015_cws} and references therein.
The developed numerical approach can be used for diluted and
concentrated random ferromagnets and it does not require a
computation of the eigenvectors of the matrix~(\ref{eq_hme});
its particular numerical efficiency for diluted systems is due to
a much smaller number of magnetic atoms as compared to the total
number of lattice sites in the supercell.

\subsection{Bulk (Ga,Mn)As and (Ga,Mn)(As,P)\label{ss_gamnas}}

\begin{figure}
\includegraphics[width=0.45\textwidth]{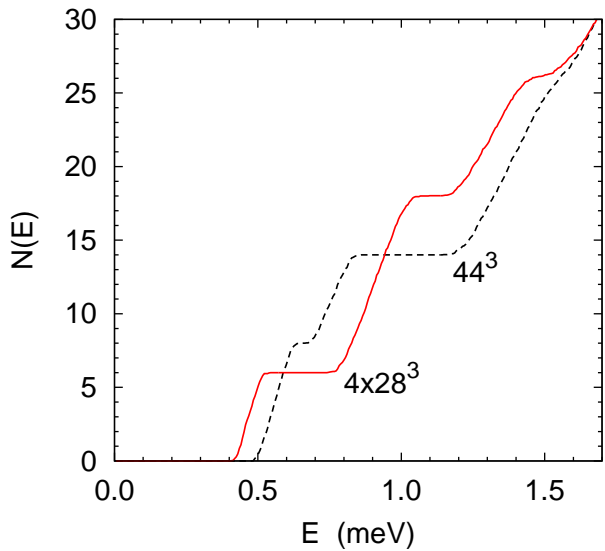}
\caption{%
The number of states $N(E)$ for GaAs doped by 5\% Mn simulated
by a cubic supercell with $4 \times 28^3$ sites (full line) and
by a rhombohedral supercell with $44^3$ sites (dashed line).
\label{f_negma}}
\end{figure}

The low-energy tail of the number of states function $N(E)$ for
GaAs doped by 5\% Mn atoms is shown in Fig.~\ref{f_negma}.
Similarly to the previous case (Fig.~\ref{f_ne2d}), plateaux
in $N(E)$ are present for both kinds of supercells.
The constant values of $N(E)$ in the first plateaux amount to six
and eight for the cubic and rhombohedral supercell, respectively.
These integer values can again be explained in terms of the shortest
reciprocal-space vectors derived from the supercell shape and size:
there are six vectors equivalent to the vector
$\mathbf{q} = 2\pi (1, 0, 0) / (La)$ for the cubic supercell, while
eight vectors equivalent to the vector $\mathbf{q} = 2\pi (1, 1, 1)
 / (La)$ correspond to the rhombohedral supercell.
Note that in the case of the rhombohedral supercell, the plateau
separating the first and the second lowest eigenvalues is quite
narrow, since the magnitude of the second shortest
$\mathbf{q}$-vector, given by $\mathbf{q} = 4\pi (1, 0, 0) / (La)$,
exceeds only slightly that of the first shortest vector.
However, a separation between the second and third shortest
vectors is clearly visible for both kinds of supercells
(Fig.~\ref{f_negma}).

\begin{figure}
\includegraphics[width=0.45\textwidth]{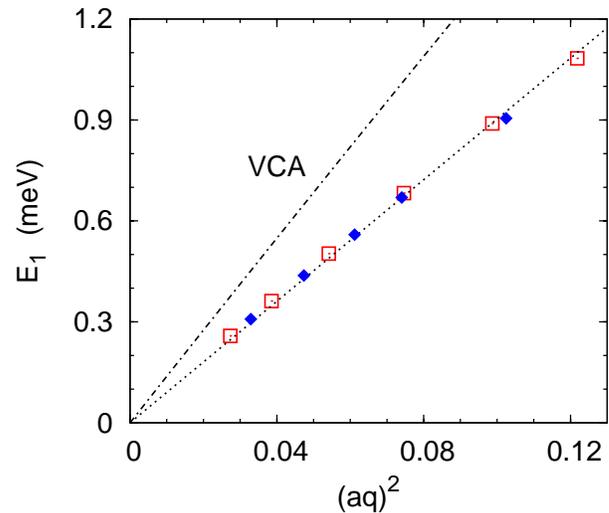}
\caption{%
The configurationally averaged
lowest excitation energies $E_1$ for Mn-doped GaAs as functions
of the $\mathbf{q}$-vector magnitude derived from the supercell
size: for cubic supercells (open boxes) and for rhombohedral
supercells (full diamonds).
The straight dotted line results from a fit and
the dash-dotted line refers to the VCA-dependence.
\label{f_eqgma}}
\end{figure}

The dependence of the average value $E_1$ of the six (cubic
supercell) or eight (rhombohedral supercell) lowest excitation
energies on the size of the shortest $\mathbf{q}$-vector,
obtained by variations of $L$ for the same system, is displayed
in Fig.~\ref{f_eqgma}.
The relation $E_1 \propto q^2$ with a slope independent on the
supercell shape is clearly visible again, in analogy with the
two-dimensional model (Fig.~\ref{f_eq2d});
this slope is markedly smaller than the value derived from the
VCA magnon dispersion law~(\ref{eq_magen_vca}).
The overestimation of the VCA with respect to the numerically
obtained slope (spin stiffness) for Mn-doped GaAs was revealed
originally by Bouzerar \cite{r_2007_gb}.

\begin{figure}
\includegraphics[width=0.45\textwidth]{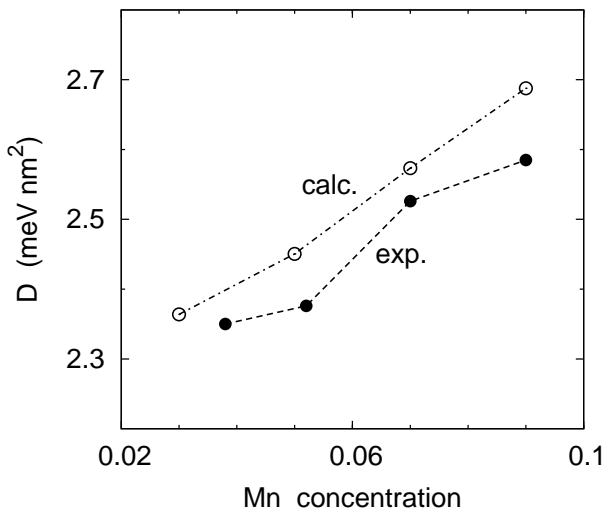}
\caption{%
The calculated spin-wave stiffness $D$ as a function of
Mn concentration in (Ga,Mn)As (open circles).
The experimental values (full circles) \cite{r_2013_nnt} are
shown as well.
\label{f_dxgma}}
\end{figure}

The calculated spin-wave stiffness $D$ of bulk (Ga$_{1-x}$Mn$_x$)As
as a function of Mn concentration $x$ is displayed in 
Fig.~\ref{f_dxgma} together with recent experimental values
obtained for epitaxial thin films with a small amount of
compensating defects \cite{r_2013_nnt}.
It should be noted that typical sizes $q$ of wave vectors relevant
in the magneto-optical pump-and-probe experiment,
which are given by the thickness of prepared epitaxial films and
by an integer index of the observed resonance mode, are
characterized by $aq \gtrsim 0.04$ (Ref.~\onlinecite{r_2013_nnt}),
which includes values smaller than those used in the numerical
simulations, $aq \gtrsim 0.17$ (Fig.~\ref{f_eqgma}).
In spite of this difference, one can see satisfactory agreement
between the theory and experiment both in the magnitude and in the
slope of the concentration dependence of $D$ (Fig.~\ref{f_dxgma}).
The present values of $D$ are quite close to the results obtained by
Bouzerar \cite{r_2007_gb}, which were based on the same
\emph{ab initio} exchange interactions (with a smaller cut-off
distance than in this study) and on the peaks of the
magnon spectral function, whereas a more recent model study,
including effects of spin-orbit coupling on the exchange
interactions, leads to appreciably smaller spin stiffnesses
\cite{r_2010_wd}.

Among other diluted ferromagnets related to (Ga,Mn)As, the
quaternary (Ga,Mn)(As,P) alloy, where P atoms substitute As atoms
on the anion sublattice of the zinc-blende structure, has recently
been investigated both theoretically \cite{r_2006_xs, r_2007_mkm}
and experimentally \cite{r_2015_srt, r_2014_tbc};
its chemical composition can be summarized as 
(Ga$_{1-x}$Mn$_x$)(As$_{1-y}$P$_y$).
The experimental study of alloys with 6\% Mn has indicated that
incorporation of 10\% of phosphorus leads to a pronounced reduction
of the spin-wave stiffness: values of $D = (2.5 \pm 0.2)$ meV~nm$^2$
and $D = (1.9 \pm 0.2)$ meV~nm$^2$ for $y=0$ and $y=0.1$,
respectively, were reported \cite{r_2014_tbc}.
However, this effect of P-doping has not been confirmed in another
study of samples of similar compositions \cite{r_2015_srt}. 

We have studied the spin-wave stiffness $D$ for $y = 0$ and
$y=0.1$ with a fixed Mn concentration, $x = 0.06$.
The P-doped system was treated with two different lattice parameters:
first, with the same $a$ as that of P-free system (identical to that
of pure GaAs, $a = 0.565$ nm) and, second, with a slightly smaller
value ($a = 0.563$ nm) which reflects the smaller lattice of pure
GaP ($a = 0.545$ nm) as compared to GaAs.
The reference value of $D$ for the P-free system ($y = 0$) was
$D = 2.52$ meV~nm$^2$, while for the P-doped system ($y = 0.1$),
values of $D = 2.48$ meV~nm$^2$ and $D = 2.47$ meV~nm$^2$ were
obtained.
The corresponding values of the Curie temperature, calculated in the
real-space RPA \cite{r_2004_hn}, are:
$T_\mathrm{C} = (124 \pm 1)$ K for $y=0$ and 
$T_\mathrm{C} = (122 \pm 1)$ K for $y=0.1$ (irrespective of $a$).
The calculated results agree thus with the observed negligible
sensitivity of $D$ to alloying by phosphorus \cite{r_2015_srt};
the origin of the reduction of $D$, reported in
Ref.~\onlinecite{r_2014_tbc}, might be ascribed to additional
compensating defects induced by P doping which should also reduce
the $T_\mathrm{C}$.
This explanation is in line with the measured Curie temperatures
\cite{r_2014_tbc}: $T_\mathrm{C} \approx 130$ K for $y=0$ and 
$T_\mathrm{C} \approx 110$ K for $y=0.1$, which
contrasts the very small change of the calculated $T_\mathrm{C}$
due to the incorporation of phosphorus, as found also in a previous
theoretical study \cite{r_2006_xs}. 

\subsection{Relation between spin-wave stiffness 
            and Curie temperature\label{ss_dtc}}

The spin-wave stiffness $D$ is related naturally to low-temperature
properties of a ferromagnet, in contrast to the Curie temperature
$T_\mathrm{C}$ as a basic finite-temperature quantity.
The mutual relation of $D$ and $T_\mathrm{C}$ has been experimentally
investigated, e.g., for Co-based Heusler alloys \cite{r_2011_uof}.
On the theoretical side, a proportionality between both quantities
can be derived; this derivation
rests on a mean-field treatment for $T_\mathrm{C}$ and on the
nearest-neighbor nature of the exchange interaction, see 
Ref.~\onlinecite{r_2010_wd} and references therein. 
Since the MFA is not accurate enough for diluted systems, where the
spatial range of $J_{mn}$ plays an important role, the relation
between the $D$ and $T_\mathrm{C}$ has been addressed in several
studies including diluted metallic Pd-based alloys \cite{r_1976_ks}
as well as the (Ga,Mn)As system \cite{r_2007_gb, r_2010_wd}.
In order to get a deeper insight into this topic, we have applied the
developed theory also to Mn-doped GaN alloys treated in the cubic
zinc-blende structure \cite{r_2004_bek, r_2004_ssd, r_2005_hn_p}.
Since the exchange interactions in (Ga,Mn)N are much more localized
than in (Ga,Mn)As, a smaller cut-off distance of $4a$ could be used
for reliable numerical values.
The Mn concentration $x$ in (Ga$_{1-x}$Mn$_x$)N was limited to
$0.02 \le x \le 0.06$ in order to avoid a negative exchange
interaction between the second nearest Mn-Mn neighbors which appears
for higher Mn contents \cite{r_2004_ssd, r_2005_hn_p}.

The values of the Curie temperature were obtained by the real-space
RPA \cite{r_2004_hn} for both systems.
The calculated $T_\mathrm{C}$'s for (Ga,Mn)As compare reasonably well
with the experiment for carefully prepared samples \cite{r_2013_nnt},
see a recent study for more details \cite{r_2016_kdt}.
The obtained $T_\mathrm{C}$'s for (Ga,Mn)N are very close to the
results of Ref.~\onlinecite{r_2005_hn_p} employing the same RPA
approach;
Monte Carlo simulations yield Curie temperatures in the studied
concentration interval both slightly above \cite{r_2004_bek} and
slightly below \cite{r_2004_ssd} the present values.

\begin{figure}
\includegraphics[width=0.45\textwidth]{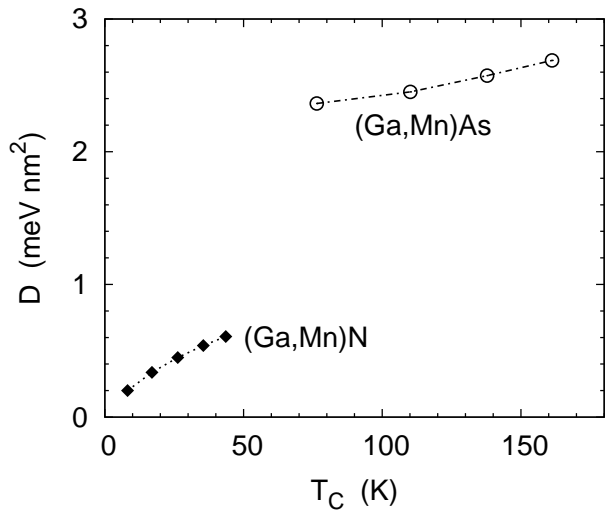}
\caption{%
Correlations between the Curie temperature $T_\mathrm{C}$
and the spin-wave stiffness $D$ as calculated for (Ga,Mn)As
(open circles) and (Ga,Mn)N (full diamonds).
\label{f_dvstc}}
\end{figure}

\begin{figure}
\includegraphics[width=0.45\textwidth]{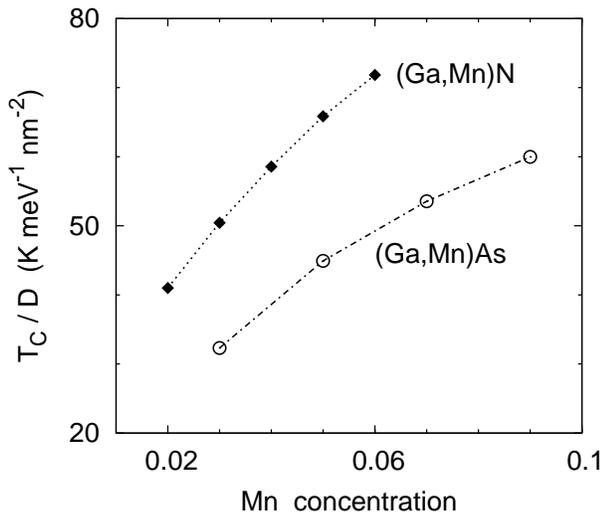}
\caption{%
The ratio $T_\mathrm{C}/D$ as a function of Mn concentration
for (Ga,Mn)As (open circles) and (Ga,Mn)N (full diamonds).
\label{f_ratio}}
\end{figure}

The resulting correlations between the $D$ and $T_\mathrm{C}$ are
displayed in Fig.~\ref{f_dvstc} for both systems.
One can see not only appreciably different values for the two
ferromagnets, but also qualitatively different trends.
However, the ratio $T_\mathrm{C}/D$, plotted in Fig.~\ref{f_ratio},
exhibits quite similar values and qualitatively identical
concentration dependences in both cases studied.
Let us note that the same concentration trend of $T_\mathrm{C}/D$,
namely, its decrease with decreasing $x$ and its concave shape,
was reported previously for (Ga,Mn)As \cite{r_2007_gb}.
The present analysis allows one to extend this conclusion also to
(Ga,Mn)N, despite the markedly stronger spatial localization of
the underlying exchange interactions.

\section{Conclusions\label{s_conc}}

We have proved that diluted ferromagnets with long-ranged exchange
interactions possess spin-wave spectra with a nonvanishing
coherent part in the low-energy region.
This coherence leads to interference effects which are not
suppressed by the strong disorder of the diluted system. 
We have shown that these features can be employed in numerical
simulations to determine the spin-wave stiffness without any
recourse to the magnon spectral function.
Our approach represents a simple and efficient way in the case of
diluted systems, that is complementary to a recent sophisticated
theory applicable also to concentrated random ferromagnets
\cite{r_2016_bsb}.

In combination with existing first-principles techniques for
calculations of the exchange interactions, we studied selected
diluted magnetic III-V semiconductors.
Besides the theoretical reproduction of the measured spin stiffness
in optimally synthesized (Ga,Mn)As samples and of its negligible
sensitivity to the doping by phosphorus, we found that the
ratio of the Curie temperature to the spin stiffness follows
qualitatively the same concentration trend independent of
the degree of localization of the Mn-Mn exchange interactions.

This pilot study was limited to highly symmetric (square, cubic)
structures and to ideal models of diluted ferromagnets within the
isotropic Heisenberg Hamiltonian. 
Consequently, straightforward extensions can be considered in the
following directions:
(i) presence of compensating defects, atomic short-range order,
or chemical inhomogeneities,
(ii) cases with less symmetric structures, such as, e.g., the
hexagonal wurtzite structure relevant for (Ga,Mn)N, and 
(iii) spin models with anisotropic exchange interactions owing
to relativistic effects.
However, possible generalization of the developed approach is less
obvious for solving more difficult problems, 
such as, e.g., a treatment of
ferromagnetic and antiferromagnetic exchange interactions, which
modify the ground state of a random magnet (a spin-glass state) and
its excitation spectrum, or investigation of the disorder-induced
damping of the spin waves (finite lifetimes of magnons), addressed
recently by means of the spectral function \cite{r_2015_cws}.
Clarification of these points has to be left for future studies.

\begin{acknowledgments}
This work was supported financially by the Czech Science
Foundation (Grant No.\ 15-13436S).
\end{acknowledgments}


\providecommand{\noopsort}[1]{}\providecommand{\singleletter}[1]{#1}%

\end{document}